\documentclass[preprint,amsmath,amssymb,superscriptaddress]{revtex4}
\usepackage{dcolumn}
\usepackage{bm}
\usepackage{graphicx}

\begin{document}

\title{The Effects of Market Properties on Portfolio Diversification in the Korean and Japanese Stock Markets
}

\author{Cheoljun Eom}
\affiliation{Division of Business Administration, Pusan National University, Busan 609-735, Republic of Korea}
\email{shunter@pusan.ac.kr}

\author{Jongwon Park}
\affiliation{Division of Business Administration, The University of Seoul, Seoul 130-743, Republic of Korea}

\author{Woo-Sung Jung}
\affiliation{Department of Physics and Basic Science Research Institute, Pohang University of Science and Technology, Pohang 790-784, Republic of Korea}
\affiliation{Center for Polymer Studies and Department of Physics, Boston University, Boston, MA 02215, USA}

\author{Taisei Kaizoji}
\affiliation{Division of Social Sciences, International Christian University, Tokyo 181-8585, Japan}

\author{Yong H. Kim}
\affiliation{College of Business, University of Cincinnati, OH 45221, USA}

\date{\today}

\begin{abstract}
In this study, we have investigated empirically the effects of market properties on the degree of diversification of investment weights among stocks in a portfolio. The weights of stocks within a portfolio were determined on the basis of Markowitz's portfolio theory. We identified that there was a negative relationship between the influence of market properties and the degree of diversification of the weights among stocks in a portfolio. Furthermore, we noted that the random matrix theory method could control the properties of correlation matrix between stocks; this may be useful in improving portfolio management for practical application.
\end{abstract}

\maketitle

\section{Introduction}
In the field of finance, portfolio management is a crucial study topic, both from an academic and a practical perspective. Markowitz's portfolio  theory has had a significant influence on portfolio research, in that it proposes a quantitative method for structuring a portfolio with a minimum (maximum) risk (return) for a given return (risk) \cite{1}. Moreover, investment weights for stocks in a portfolio require well-distributed diversification to effectively reduce the risk of a portfolio. Diversification is critical for the application of Markowitz's optimum portfolio from a practical viewpoint, and the two factors that can affect diversification are the number of stocks in a portfolio and the level of distribution of the investment weights.

First, previous literature concerning the effects of the changes in the number of stocks within a portfolio on reducing the risk by diversification have provided relatively robust empirical evidences. Evans \textit{et al.} previously suggested that a portfolio's risk can be reduced by 50\% or more if the number of stocks in a portfolio is $10 \sim 15$, and also provided empirical evidence suggesting that the addition of more stocks into the portfolio exerts only insignificant risk reduction effects ($2\sim 5\%$) \cite{2}. Without question, the risk reduced by diversification is an unsystematic risk associated with stocks \cite{3}. Their research findings provide important criteria for determining the number of stocks in practical portfolio management. Elton \textit{et al.} presented an analytical equation which allows for an examination of the degree of risk reduction when adding new stocks to a portfolio \cite{4}. Statman assessed the effects of adding stocks to a portfolio on reducing the risk of an existing portfolio from the standpoints of marginal benefits obtainable from reducing the portfolio risk and the marginal costs that occur as the result of increasing the number of stocks in a portfolio \cite{5}. The study suggested that, for practical purposes, the number of stocks should be increased to 30 (for a borrowing investor) or 40 (for a lending investor) in order to realize the effects of portfolio diversification.

Next, there are studies that address the level of distribution of investment weights among stocks in a portfolio, which is crucial to the implementation of an effective portfolio diversification. These studies attempted to determine whether the total investment is distributed well among stocks within a portfolio. However, on the basis of the empirical evidence, although an efficient portfolio with minimized risk for a given return from traditional portfolio theory is created, the investment weights among stocks within a created portfolio are concentrated on a few stocks--that is to say, they are not well distributed. Such investment weights of lower diversification among stocks within a portfolio function as a limitation to the use of the portfolio theory in practical applications.

According to the random matrix theory (RMT) recently derived from the field of econophysics to the field of finance \cite{6,7}, observed results exist which indicate that portfolio management can be improved by controlling the correlation matrix, as an important input data in the Markowitz's optimum portfolio. Financial time series data contains noise, finiteness of time series, missing data, and thin trading - these factors force the correlation matrix calculated from the actual data to contain measurement errors. Therefore, the RMT can improve the portfolio theory by removing the measurement error inherent to the correlation matrix. According to the results obtained by combining the RMT with Markowitz's optimum portfolio \cite{8,9,10}, using the correlation matrix that controlled the correlation matrix of a past period on the basis of the RMT method, as the estimated correlation matrix of a future period, provides a better approximation of the portfolio elicited from the actual correlation matrix of the future period than the conventional approach using a correlation matrix of a past period as the estimated correlation matrix of a future investment period. The results also show that the utility of the RMT improves reliability in portfolio selection, as well as accuracy and stability in portfolio risk evaluation.

The primary objective of our study is to examine empirically whether the level of distribution of investment weights for stocks in a portfolio can be improved by controlling the divere properties included in the correlation matrix. Specifically, we believe that among the various properties in the correlation matrix, the market factor properties exert critical effects on the degree of diversification of investment weights in Markowitz's optimum portfolio. Accordingly, we assessed the effects of changes in the influence of market factor properties on the degree of diversification of investment weights among stocks in a portfolio.

First of all, we selected market factor as the properties that principally influence the correlation among stocks that constitute a portfolio. Many models developed in the field of finance to elucidate the pricing mechanism of the stock market commonly utilize market factor as an explanation variable. King previously provided empirical evidence suggesting that stock price changes can be explained using market, industry, and company factors and market factor, in particular, account for a significant portion of stock price changes in the market \cite{11}. These observed results affected the CAPM (capital asset pricing model) \cite{12} and APM (arbitrage pricing model) \cite{13}. In addition, studies that have applied the RMT method to the field of finance determined that the largest eigenvalue has the properties of market factor \cite{8,9,10}, and similar results have been noted in other financial research \cite{11,14,15}. The eigenvector is utilized as the weight for creating time series data (corresponds to the factor score in the multivariate statistics of principal component analysis \cite{17}) that reflects the eigenvalue properties. The eigenvector distribution of the largest eigenvalue has a higher average than the eigenvector distributions of other eigenvalues \cite{10}. Moreover, it has been verified that the properties of the largest eigenvalue do not change, regardless of the number and type of stocks in a correlation matrix \cite{16}. Among the properties included in the correlation matrix, we conducted an empirical investigation of the effects of market factor properties, in which defined by the largest eigenvalue via the RMT, on portfolio diversification, in which we comparatively examined the degree of diversification of the investment weights among stocks in a portfolio by categorizing them into cases with and without market factor properties in the correlation matrix.

Next, previous studies have reported that market factor exerts significant effects on the stock market during a market crisis. King provided empirical evidence suggesting that the degree of price fluctuation of stocks that could be explained by market factor (using the $R^2$ determinant coefficient) was relatively high during the Great Depression of the 1920s \cite{11}. According to a study conducted by Onnela \textit{et al.}, the magnitude of the correlation among stocks has increased in times of market crisis, such as "Black Monday" in the U.S. in October 1987 \cite{18,19}. Eom \textit{et al.} noted that the frequency of significant information flow among stocks increased rapidly in Korea during the Asian foreign exchange crisis of 1997 \cite{20}. Therefore, we have empirically assessed changes occurring in the influence of market factor in Korean and Japanese stock markets in and around the Asian foreign exchange crisis in December 1997, and the large Japanese recession in January 1990, respectively. We further attempted to determine the manner in which investment weight distributions among stocks in a portfolio were affected by market factor during market crisis situations.

In summary, our primary objective was to investigate empirically the usefulness of the RMT as a potential method for improving the degree of diversification of the stocks in a portfolio constructed in accordance with conventional portfolio theory, by controlling market factor among the properties included in the original correlation matrix. On the basis of the observed results, we were able to confirm the empirical evidence suggesting that the influence of market factor increases during a market crisis. We also noted that the distribution level of the investment weights among stocks in a portfolio constructed from the Markowitz's optimum portfolio is reduced when market factor exerts a significant influence. However, the distribution level of investment weights for stocks in a portfolio elicited from a correlation matrix without market factor properties by using the RMT method was insensitive to changes in the effects of market factor. Moreover, the risk of a portfolio elicited from a correlation matrix without market factor properties was smaller than that of a portfolio drawn from the conventional Markowitz's optimum portfolio. These results indicate that in a stock market, the market factor properties evidence a negative relationship with the distribution level of investment weights of stocks in a portfolio. We also learned that the combination of a correlation matrix controlled by the RMT method and the portfolio theory is useful for improving portfolio management in practical terms.

This paper is constructed as follows. The first chapter is the introduction and Chapter 2 describes the data and methods employed in this study. Chapter 3 presents the results observed according to the study objectives established previously. The final chapter offers a summary of observed results and their implications.

\section{Data and methods}
\subsection{Data}
We utilized daily price data of KOSPI 200 stocks in the Korean stock market and Nikkei 225 stocks in the Japanese stock market. Among the stocks included in the market indices, we determined the ones to be utilized for the process on the basis of the following three criteria. First, stocks with consecutive daily stock prices for the period were selected. Second, stocks with statistical extremes in terms of skewness ($>|2|$) and kurtosis ($>30$) were excluded. Third, stocks in sectors with four or less companies were excluded. In turn, we selected 104 stocks from Korea's KOSPI200 and 183 from Japan's Nikkei225 that fulfilled the three criteria.

According to the objective of our study, we took the price data during the periods that included market crises when the influence of market factor in the stock markets increased. For the Korean stock market, this was between January 1990 to December 2007 (216 months), a period that includes the foreign exchange crisis of December 1997. For the Japanese stock market, we compiled data from January 1983 to December 2000 (216 months), a duration which includes the beginning of the large recession in January 1990. We utilized the rolling sample method to assess changes in the effects of market factor, and to ensure variation in the degree of diversification of investment weights of stocks within a portfolio from the perspective of the time series. For the entire period (216 months), we established a testing period duration of 5 years (60 months) and 1 month of duration shift. Ultimately, we noted the time series variation in the results obtained from 157(=216-60+1) repetitions for the Korean and Japanese stock markets, respectively.

\subsection{Methods}
The objective of this study was to investigate empirically the effects of changes in the influence of market factor properties in the stock market on the distribution level of investment weights among stocks in a portfolio. Therefore, we required methods to measure quantitatively the degree of influence of market factor and the degree of diversification of investment weights for stocks in a portfolio.

Initially, we utilized two parameters for the quantitative measurement of the degree of influence of market factor. First, the largest eigenvalue, $\lambda_1$ was elicited via the RMT method, and second, the mean square error (MSE) of the difference between the original correlation matrix, $C^O$ with the properties of market factor and the controlled correlation matrix, $C^M$ without the market factor properties. We employed the RMT method to control the market factor included in the correlation matrix among stocks.

The first measurement used to quantify the degree of influence of market factor is the largest eigenvalue. By the statistical properties of the correlation matrix created by the random matrix, if the length of time series, $L$, and the number of data, $N$, is infinite, the probability density function $P_{RM}(\lambda)$ of the random correlation matrix eigenvalue, $\lambda$ can be defined analytically as follows[21].

\begin{eqnarray}
P_{RM}(\lambda)&=&\frac{Q}{2\pi}\frac{\sqrt{(\lambda_{+}^{RM}-\lambda)(\lambda-\lambda_{-}^{RM})}}{\lambda}\\
( \lambda_{\pm}^{RM}&=&1+\frac{1}{Q}\pm 2 \sqrt{\frac{1}{Q}},~Q\equiv\frac{L}{N}>1 ) \nonumber
\end{eqnarray}

\noindent and the range of the eigenvalue included in a random matrix is $\lambda_{-}^{RM}\leq\lambda_i\leq\lambda_{+}^{RM}$, where $\lambda_{+}^{RM}$ and $\lambda_{-}^{RM}$ denote the maximum and minimum eigenvalues, respectively.

From previous studies, it is widely recognized that the largest eigenvalue among eigenvalues that exceed the range of a random matrix, $\lambda_i>\lambda_{+}^{RM}$ evidences market factor properties \cite{6,7,8,9,10,11,14,15,16}. We utilized the largest eigenvalue as the proxy for market factor, and noted the time series variation of its magnitude. In other words, an increasing magnitude of the largest eigenvalue is indicative of a greater market factor influence in the stock market, and vice versa.

The second measurement is the MSE. Based on previous studies, we generated a correlation matrix, $C^M$, from which the properties of the largest eigenvalue were removed via the RMT method \cite{22}. This is the correlation matrix without market factor properties. The MSE is assessed using the off-diagonal elements $N(N-1)/2$ of the controlled correlation matrix, $C^M$, and the original correlation matrix, $C^O$, as follows.

\begin{equation}
MSE=\sqrt{\frac{1}{N(N-1)/2} \sum_{k=1}^{N(N-1)/2} (C_k^O - C_k^M)^2}
\end{equation}

\noindent in which a high MSE reflects a high influence from market factor because the MSE is the difference between $C^O$ with market factor and $C^M$ without market factor. If the influence of market factor increases, $C^M$ will decrease, causing the MSE to assume a large value. On the other hand, a small MSE is reflective of an insignificant market factor influence.

Next, with regard to the quantitative measurement of the degree of diversification of investment weights among stocks in a portfolio, we employed two parameters: the intra-portfolio correlation ($IPC$) and the concentration coefficient ($CC$). These parameters were calculated using the investment weights of stocks constituting a portfolio. Accordingly, we utilized a method predicated on Markowitz's portfolio theory to generate investment weights for stocks in a portfolio.

\begin{eqnarray}
\sigma_p&=&\sqrt{\sum_i \sum_j w_i^p w_j^p \sigma_{i,j}} ~~~ (\sigma_{i,j}=\sigma_i \cdot \sigma_j \cdot \rho_{i,j})\\ \nonumber
\textrm{CONDITION }1&:&E(R_p)=\sum_j w_j^p E(R_j)=R_p^T ~~ (p=1,2,\dots,10) \\ \nonumber
\textrm{CONDITION }2&:&\sum_j w_j^p\equiv1.0 \\ \nonumber
\textrm{CONDITION }3&:&w_j^p \geq 0.0 ~~ (j=1,2,\dots,N) \nonumber
\end{eqnarray}

\noindent where $\sigma_i$ and $\sigma_j$ are stock risks (standard deviation), and $\rho_{i,j}$ ($=C^O$) denotes the correlation among stocks. Condition 1, which is concerned with the expected return of portfolio, $E(R_p)$, specifies that expected returns, $E(R_j)$ of stocks in a portfolio have a certain target return, $R_p^T$. Then, a portfolio is generated with a minimum risk, $\sigma_p$ according to the minimization objective function of Eq. 3 for the target return; the investment weight, $w_j^p$ for stocks in a portfolio are created in the process. Finally, connecting the combination points $[\sigma_p,~E(R_p)]$ of portfolio risks and returns created by varying the target return within a range, $p=1,2,\dots,10$ provides an efficient investment curve, or an efficient portfolio. Conditions 2 and 3 indicate that short-selling is not permitted.

In order to obtain robust results, we conducted an identical testing process using the controlled correlation matrix from which the market factor, $\lambda_1$ were removed via the RMT method, as the correlation matrix input data $\rho_{i,j}\rightarrow C^M$ of Eq. 3. For an objective comparison between portfolios elicited from correlation matrices with the market factor properties and without the market factor properties, we set an identical target return for both cases.

Using investment weights for stocks in a portfolio calculated from the optimization function of Eq. 3, we attempted to determine whether investment weights were well distributed among stocks with the $IPC$ and the $CC$. The first measurement $IPC_p$ quantifies the distribution level of the investment weights among stocks in a portfolio.

\begin{equation}
IPC_p=\frac{\sum_i^N \sum_j^N w_i^p w_j^p \rho_{i,j}}{\sum_i^N \sum_j^N w_i^p w_j^p} ~~ (p=1,2,\dots,10)
\end{equation}

\noindent The range of the IPC is $-1\leq IPC \leq 1$. $IPC=-1$ indicates that the investment weights are well distributed among all of stocks comprising the portfolio, and $IPC=+1$ shows that the investment weights are not distributed at all. Namely, the lower the IPC, the higher the level of distribution for the investment weights among stocks in a portfolio.

The second measurement, $CC_p$, as a complementary parameter to $IPC_p$, measures the degree of concentration of investment weights among stocks in a portfolio.

\begin{equation}
CC_p=(\sum_{j=1}^N (w_j^p)^2)^{-1} ~~ (p=1,2,\dots,10)
\end{equation}

\noindent The range of the CC is $1\leq CC\leq N$. $CC=1$ denotes that 100\% of the investment is made on one stock in a portfolio, and $CC=N$ means that an equal investment weight $w_j= \frac{1}{N} =w$ is applied to every stock in the portfolio. In other words, the higher the $CC$, the lower the degree of concentration of investment weights among stocks in a portfolio.

\section{Results}
This section presents the results of our examination of the effects of market factor properties on the degree of diversification of investment weights among stocks in a portfolio. The results are divided into those obtained from the Korean stock markets and those obtained from Japanese stock markets, as is shown in Fig. 1 and Fig. 2, respectively. The results were generated through 157(=216-60+1) repetitions (based on the rolling sample method) for a testing period of 5 years and 1 month of duration shift, for a total of 18 years. Figs. 1 \& 2 (a) depict the trends in the market indices over the total period. We included Korea's foreign exchange crisis in December 1997 and the beginning of Japan's big recession in January 1990 during this period. Figs. 1 \& 2 (b) show the results of quantitative measurements of the influence of market factor in the stock market. Via the RMT method, variation in the largest eigenvalue ($\lambda_1$, blue squares) and the MSE (red circles) were measured as the difference between the controlled correlation matrix without market factor properties and the original correlation matrix with market properties are indicated. Figs. 1 \& 2 (c) \& (e) show the $\overline{IPC}$ and $\overline{CC}$, which measure the degree of diversification of investment weights among stocks in portfolios. According to the varying target returns, $R_p^T$, we applied the calculated investment weight $w_j^p$ to Eqs. 4 and 5 to measure $IPC_p$ and $CC_p$, and their averages $\overline{IPC}=\frac{1}{10}\sum_{p=1}^{10} IPC_p$ and $\overline{CC}=\frac{1}{10}\sum_{p=1}^{10}CC_p$ are shown in the graphs. These results are observed for the original correlation matrix with the market factor properties (red circles) and the controlled correlation matrix without market factor properties (blue squares). In Figs. 1 \& 2 (d) \& (f), the fluctuation ranges of $\overline{IPC}$ and $\overline{CC}$ during the total period are displayed in box-plots.

According to the observed results, we determined that the distribution level of investment weights for stocks in a portfolio according to the conventional portfolio theory is sensitive to changes in the influence of market factor, and evidences a negative relationship. Figs. 1 \& 2 (a) \& (b) demonstrate that market is volatile and time-varying, and the two parameters ($\lambda_1$, MSE) indicative of the influence of market factor increase rapidly during market crises. These results constitute empirical evidence that the effects of market factor are increased during times of market crisis. In Figs. 1 \& 2 (c) \& (e), during market crises when the influence of market factor increased, $\overline{IPC}$ (red circles in Figs (c)) evidenced a clear upward trend, whereas $\overline{CC}$ (red circles in Figs (d)) evidenced a decreasing trend. Namely, the degree of diversification of investment weights for the stocks using the original correlation matrix with the properties of market factor was quite low. On the other hand, when the distribution level of investment weights for the stocks in a portfolio was elicited using the correlation matrix without market factor properties via the RMT method ($\overline{IPC}$ and $\overline{CC}$, blue squares), no notable trend was observed. Moreover, in Figs. 1 \& 2 (d) \& (f), the magnitude of $\overline{IPC}$ ($\overline{CC}$) calculated with RMT for the entire period was significantly less (greater) than that of $\overline{IPC}$ ($\overline{CC}$) calculated in accordance with the conventional portfolio theory. Also, we represent average value of $\overline{IPC}$ and $\overline{CC}$ in Table 1, in order to summary observed results in Figs. 1 \& 2. From the observed results, we found that the degree of diversification of investment weights for stocks from using the controlled correlation matrix without the market factor properties was relatively high. Furthermore, we were able to confirm empirically that the RMT method, which can control various properties included in a correlation matrix, is an effective means of improving the degree of diversification for investment weights among the stocks in a portfolio constructed via the conventional portfolio theory.

In order to provide robust empirical evidence for the observed results, we presented the relationships among the largest eigenvalue $\lambda_1$ [Fig. 2(b)], $\overline{IPC}$ [Fig. 2(c)], and $\overline{CC}$ [Fig. 2(e)] from the results of the Japanese stock market data in Fig. 3. In Fig. 3 (a) \& (b), the largest eigenvalue displays correlations of significant 96.58\% and significant -73.84\% with $\overline{IPC}$ and $\overline{CC}$ at the 1 \% level, respectively. It can be noted that as the magnitude of the largest eigenvalue that indicates the influence of market factor increases, the level of distribution of the investment weights for stocks in a portfolio clearly decreases ($\overline{IPC}$ increases) and the concentration level of investment weights increases ($\overline{CC}$ decreases). In addition to these results, we observed fluctuation of portfolio risks in Fig. 4. Fig. 4 (a) and (b) are depicted using data from the Korean and Japanese stock markets, respectively. In figure, we also determined that the portfolio risk elicited from the controlled correlation matrix without market factor ($\overline{\sigma_p^M}=\frac{1}{10}\sum_{p=1}^{10}\sigma_p$, blue squares) was decidedly lower than the portfolio risk drawn from the conventional portfolio theory ($\overline{\sigma_p^O}$, red circles) for a given return over the total period. That is to say, average value of $\overline{\sigma_p^M}$ is 0.0033 (t:43.08) within range $ 0.0081 \leq  \overline{\sigma_p^M} \leq 0.0055 $ for data of Korean stock market, and 0.0018 (t:37.67) within range $ 0.0010 \leq  \overline{\sigma_p^M} \leq 0.0032 $ for Japanese stock market. Otherwise, average value of $\overline{\sigma_p^O}$ is 0.0131 (t:50.27)) within range $ 0.0073 \leq  \overline{\sigma_p^O} \leq 0.0180 $ for data of Korean stock market, and 0.0089 (t:66.23) within range $ 0.0067 \leq  \overline{\sigma_p^O} \leq 0.0123 $ for Japanese stock market. Therefore we robustly discovered that the properties of market factor causes to be lower degree of diversification, and then to increase the degree of portfolio risk. Also, these results show that the RMT method, which can control market factor from a correlation matrix, is an effective means not only for improving the degree of diversification for investment weights among stocks in a portfolio, which has been a practical limitation to the application of conventional portfolio theory, but also may prove useful in reducing the risk level of a portfolio.

\section{Conclusions}
This study comprised our empirical investigation of a method to improve the degree of diversification of investment weights for stocks that constitute a portfolio in the Korean and Japanese stock markets. An empirical test was conducted in order to determine whether the distribution level of investment weights for stocks in a portfolio elicited in accordance with the conventional Markowitz's portfolio theory could be improved by removing the market factor properties included in the correlation matrix via the RMT method. The observed results can be summarized as follows. We were able to confirm that the influence of market factor increased during market crises. We also determined that when the influence of market factor was high, the distribution level of investment weights for stocks in a portfolio drawn from the Markowitz's optimum portfolio decreased. In other words, we noted a negative relationship between the influence of market factor and the distribution level of investment weights for stocks in a portfolio. In order to acquire more robust results, we conducted an identical testing process using the controlled correlation matrix from which market factor were eliminated using the RMT method. The results revealed no significant variations in terms of the distribution level of investment weights for the stocks in a portfolio during times of market crisis. Moreover, compared to the results observed from the conventional portfolio theory, the results acquired via the RMT method evidenced a significantly higher degree of diversification of investment weights for stocks in a portfolio, and also evidenced a lower portfolio risk. From these results, we evidence that the properties of market factor cause to decrease the degree of diversification of investment weights among stocks in a portfolio and to increase portfolio risks. Also, we were able to verify that the RMT method that can control various properties from the correlation matrix, is an effective means of improving the practical limitations to the application of the conventional portfolio theory. We also learned that there is a clear necessity for more profound research efforts in the future regarding the roles of market factor in the portfolio theory.

\begin{acknowledgements}
This work was supported by the Korea Science and Engineering Foundation (KOSEF) grant funded by the Korea government (MEST) (No. R01-2008-000-21065-0)
\end{acknowledgements}

\newpage
\clearpage

\begin{figure}
\includegraphics[width=1.0\textwidth]{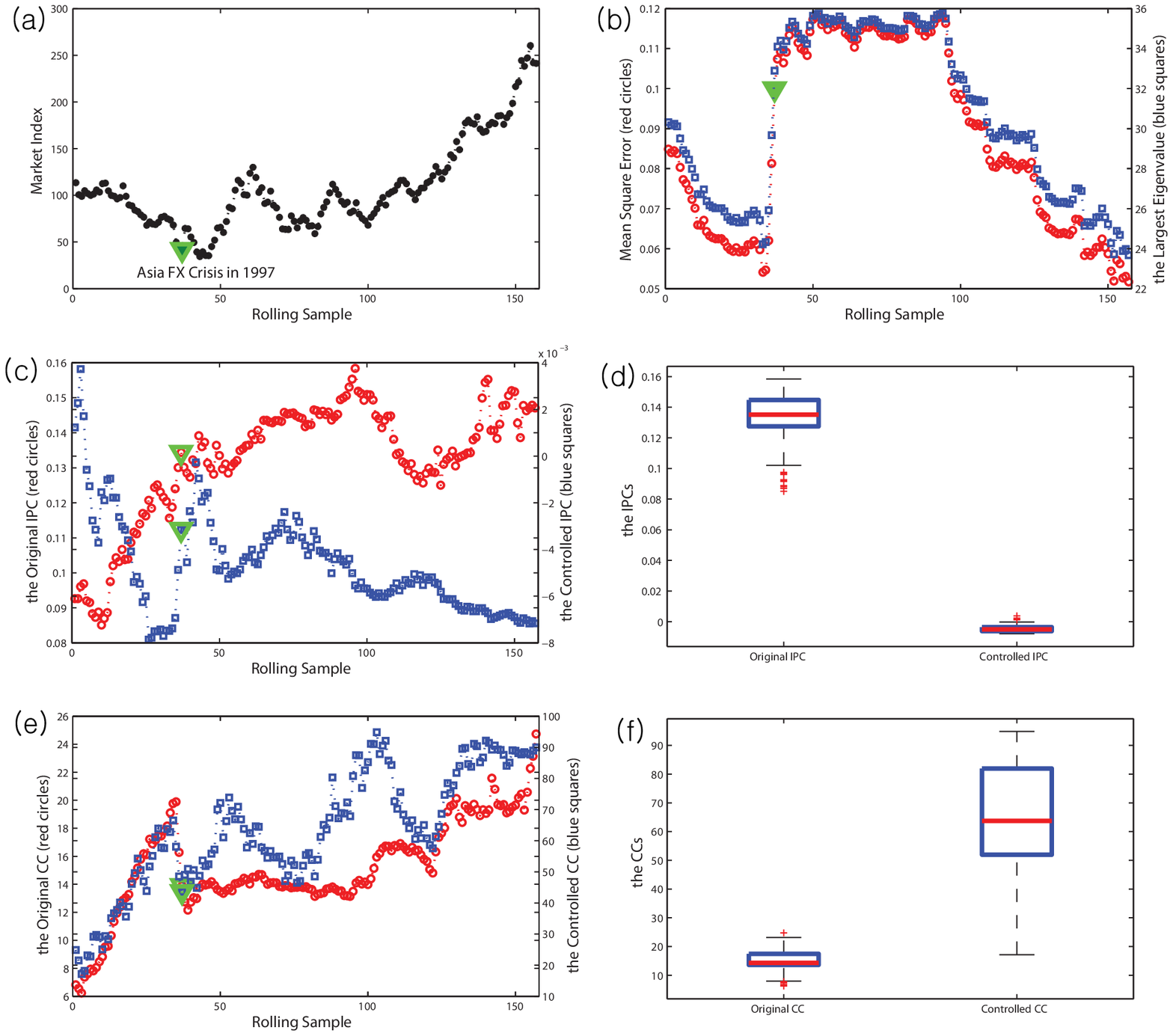}
\caption{This shows the results from Korean stock markets. Fig. 1(a) depicts the market index trends, during the period including Korea's foreign exchange crisis of December 1997. Fig. 1(b) displays the results of quantitative measurements of the influence of market factor in the stock market, using the largest eigenvalue(blue squares) and the MSE (red circles). Fig. 1 (c) \& (e) displays the degree of diversification of investment weights among stocks in portfolios using $\overline{IPC}$ and $\overline{CC}$. These results are provided for the original correlation matrix (red circles) and the controlled correlation matrix (blue squares). In Fig. 1 (d) \& (f), the fluctuation ranges of $\overline{IPC}$ and $\overline{CC}$ during the total period are displayed in box-plots.}
\label{fig:1}
\end{figure}

\newpage
\clearpage

\begin{figure}
\includegraphics[width=1.0\textwidth]{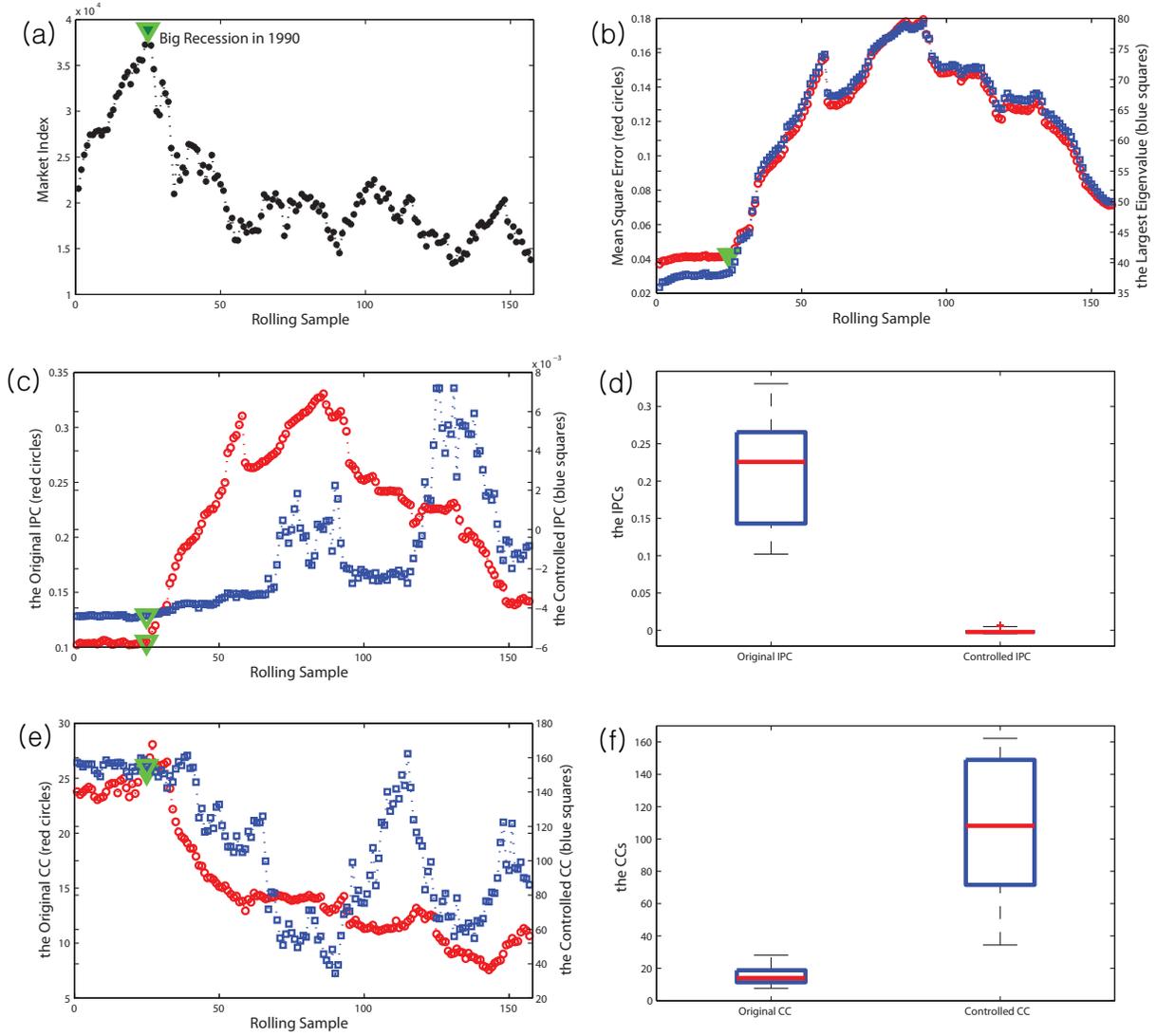}
\caption{This presents the results obtained from Japanese stock markets. Fig. 2(a) depicts the market index trends during the period including Japan's big recession of January 1990. Fig. 2(b) displays the results of quantitative measurements of the influence of market factor in the stock market, using the largest eigenvalue (blue squares) and the MSE (red circles). Fig. 2 (c) \& (e) displays the degree of diversification of investment weights among stocks in portfolios. These results are provided for the original correlation matrix (red circles) and the controlled correlation matrix (blue squares). In Fig. 2 (d) \& (f), the fluctuation ranges of $\overline{IPC}$ and $\overline{CC}$ during the total period are displayed in box-plots.}
\label{fig:2}
\end{figure}

\newpage
\clearpage

\begin{figure}
\includegraphics[width=1.0\textwidth]{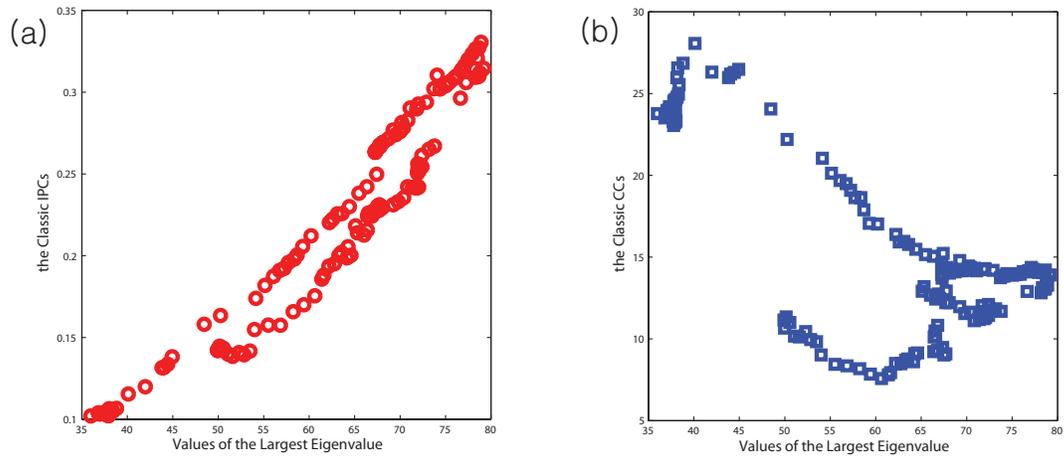}
\caption{This shows the relationships among the largest eigenvalue from the results of the Japanese stock market data in the largest eigenvalue, $\overline{IPC}$, and $\overline{CC}$. In Fig. 3, the X-axis denotes the largest eigenvalue, and the Y-axis represents $\overline{IPC}$ (Fig. 3(a)) and $\overline{CC}$ (Fig. 3(b)).}
\label{fig:3}
\end{figure}

\newpage
\clearpage

\begin{figure}
\includegraphics[width=1.0\textwidth]{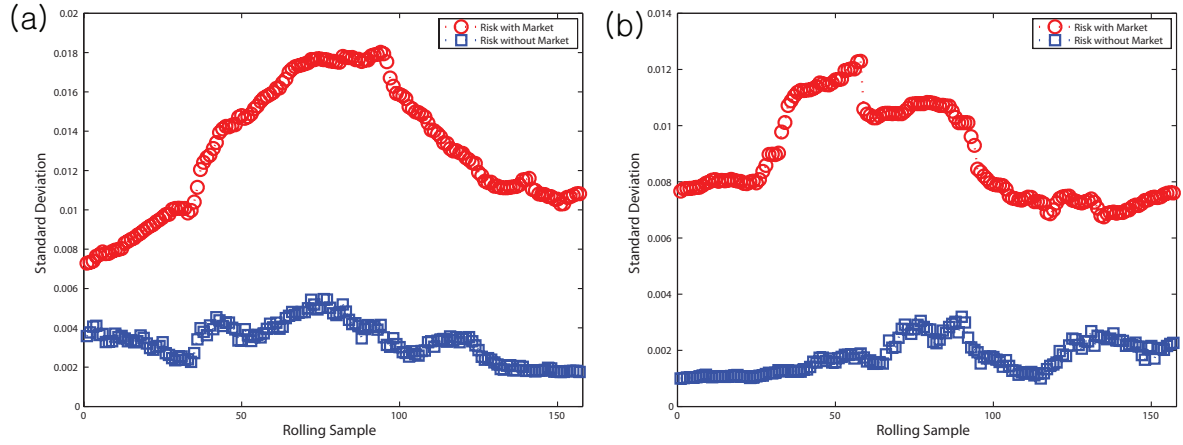}
\caption{This shows the results of comparison between portfolio risks from correlation matrices with (blue squares) and without (red circles) the market factor properties. Fig. 4 (a) and (b) are depicted using the data from the Korean and Japanese stock markets, respectively. }
\label{fig:4}
\end{figure}

\newpage
\clearpage

\begin{table}
\begin{tabular}{|c|c|c|c|}
\hline
\multicolumn{2}{|r|}{Correlation Matrices}&Original&Controlled\\
\multicolumn{2}{|l|}{Measurements}&Correlation Matrix&Correlation Matrix\\
\hline
Korean&Average of $\overline{\textrm{IPC}}$&0.1322$^*$&-0.0047$^*$\\ \cline{2-4}
Market&Average of $\overline{\textrm{CC}}$&15.24$^*$&63.68$^*$\\
\hline
Japanese&Average of $\overline{\textrm{IPC}}$&0.2120$^*$&-0.0015$^*$\\ \cline{2-4}
Market&Average of $\overline{\textrm{CC}}$&15.30$^*$&107.03$^*$\\
\hline
\end{tabular}
\caption{This table represents the results of the average value of $\overline{\textrm{IPC}}$ and $\overline{\textrm{CC}}$ during rolling sample period. ($*$: significant at the 1\% level)}
\label{fig:4}
\end{table}


\begin{thebibliography}{0}
\bibitem{1}
H. Markowitz, ``Portfolio Selection," the Journal of Finance 7(1), 1952, 77-91.

\bibitem{2}
J. L. Evans and S. H. Archer, ``Diversification and the Reduction of Dispersion: An Empirical Analysis," the Journal of Finance 23(5), 1968, 761-767.

\bibitem{3}
W. F. Sharpe, ``A Simplied Model for Portfolio Analysis," Management Science 9(2), 1963, 277-293.

\bibitem{4}
E. J. Elton and M. J. Gruber, ``Risk Reduction and Portfolio Size: An Analytical Solution," the Journal of Business 50(4), 1977, 415-437.

\bibitem{5}
M. Statman, ``How Many Stocks Make a Diversified Portfolio," the Journal of Financial and Quantitative Analysis 22(3), 1987, 353-363.

\bibitem{6}
L. Laloux, P. Cizeau, J. Bouchaud, and M. Potters, ``Noise Dressing of Financial Correlation Matrices," Physical Review Letters 83(7), 1999, 1467-1470.

\bibitem{7}
V. Plerou, P. Gopikrishnan, B. Rosenow, L. A. N. Amaral, and H. E. Stanley, ``Universal and Nonuniversal Properties of Cross Correlations in Financial Time Series," Physical Review Letters 83(7), 1999, 1471-1474.

\bibitem{8}
L. Laloux, P. Cizeau, M. Potters, and J.-P. Bouchaud, ``Random Matrix Theory and Financial Correlations," International Journal of Theoretical and Applied Finance 3(3), 2000, 391-397.

\bibitem{9}
B. Rosenow, V. Plerou, P. Gopikrishnan, and H. E. Stanley, ``Portfolio optimization and the random magnet problem," Europhysics Letters 59(4), 2002, 500-506.

\bibitem{10}
S. Sharifi, M. Crane, A. Shamaie, and H. Rukin, ``Random Matrix Theory for Portfolio Optimization: a stability approach," Physica A 335, 2004, 629-643.

\bibitem{11}
B. F. King, ``Market and Industry Factors in Stock Price Behavior," the Journal of Business 39(1),  1966, 139-190.

\bibitem{12}
F. Black, M. Jensen, and M. Scholes, ``The Capitial Asset Pricing Model: Some Empirical Tests," Working paper from SSR, 1972.

\bibitem{13}
S. A. Ross, ``The Arbitrage Theory of Capital Asset Pricing," Journal of Economic Theory 13, 1976, 343-362.

\bibitem{14}
C. Trzcinka, ``On the Number of Factors in the Arbitrage Pricing Model," the Journal of Finance 41(2), 1986, 347-368.

\bibitem{15}
S. T. Brown, ``The Number of Factors in Security Returns," the Journal of Finance 44(5), 1989, 1247-1262.

\bibitem{16}
C. Eom, W.-S. Jung, T. Kaizoji, and S. Kim, ``Effect on Eigenvalue by Changing Sample Size in the Korean and Japanese Stock Markets," preprint, available on web, arXiv.org:0811.4021, 2008.

\bibitem{17}
H. H. Harman (1976), Modern Factor Analysis, The University of Chicago Press.

\bibitem{18}
J.-P. Onnela, A. Chakraborti, K. Kaski, and J. Kertesz, ``Dynamic Asset Trees and Black Monday," Physica A 324, 2003, 247-252.

\bibitem{19}
J.-P. Onnela, A. Chakraborti, K. Kaski, J. Kertesz, and A. Kanto, ``Dynamics of Market Correlations : Taxonomy and Portfolio Analysis," Physical Review E 68, 2003, 056110.

\bibitem{20}
C. Eom, O. Kwon, and W.-S. Jung, ``Statistical Properties of Information Flow in Financial Time Series," preprint, available on web, arXiv.org:0811.0448, 2008.

\bibitem{21}
A. M. Sengupta, and P. P. Mitra, ``Distributions of Singular Values for some Random Matrices," Physical Review E 60, 1999, 389-392.

\bibitem{22}
C. Eom, G. Oh, W.-S. Jung, H. Jeong, and S. Kim, ``Topological Properties of Stock Networks based on Minimal Spanning Tree and Random Matrix Theory in Financial Time Series," Physica A 388, 2009, 900-926.

\end{thebibliography}
\end{document}